\documentclass[doublecol,graphicx,times]{epl2} 
\usepackage{graphicx}
\usepackage{amsmath}
\usepackage{amsfonts}
\usepackage{amssymb}
\usepackage{amsbsy}

\def\ag{{Ann.\ Geo\-phy\-si\-c\ae}}
\def\asr{{Adv.\ Space Res.}}

\def\grl{{Geo\-phys.\ Res.\ Lett.}}

\def\jgr{{J.\ Geo\-phys. Res.}}

\def\prl{{Phys.\ Rev.\ Lett.}}

\def\pf{{Phys. Fluids}}

\def\prl{{Phys. Rev. Lett.}}

\title{On the competition between the mirror and electromagnetic ion cyclotron modes}
\shorttitle{Mirror mode} 

\author{R. A. Treumann\thanks{Present address: International Space Science Institute, Bern, Switzerland, e-mail: treumann@issibern.ch}\inst{1,2} \and O. D. Constantinescu$^3$}
\shortauthor{R. A. Treumann and O. D. Constantinescu}

\institute{ 
  \inst{1}Department of Geophysics, Munich University, Theresienstr. 41, D-80333 Munich, Germany\\                   
  \inst{2}Department of Physics and Astronomy, Dartmouth College, Hanover, NH 03755\\
  \inst{3}Institute of Geophysics and extraterrestrial Physics, Technical University Braunschweig, Braunschweig, Germany  
}
\pacs{96.50.Pw}{Particle acceleration}
\pacs{96.50.Vg}{Energetic particles}
\pacs{94.05.Lk}{Turbulence}

\abstract{
We give a simple argument for the exclusive existence of mirror and electromaghetic ion cyclotron modes in anisotropic high-$\beta$ plasmas. It is shown that, in addition to a large domain of coexistence of both modes, two domains exist in parameter space $(A,\beta_\perp)$ where solely either mirror modes or electromagnetic ion cyclotron modes can be excited. In the overlap region the modes with the larger growth rate should win. However nonlinear effects may modify such a conclusion.}

\begin{document}

\maketitle
\section{\textsf{Introduction}}
The mirror instability is one of the fundamental fluid modes in an anisotropic plasma \cite{chandra1958,hase1969} hosting an excess of perpendicular energy in the form of  enhanced perpendicular pressure $P_\perp>P_\|$. The complementary instability is the firehose mode which is excited when the above pressure condition is inverted. One conveniently defines a pressure (or temperature) anisotropy $A=(P_\perp/P_\|)-1$ which quantifies the perpendicular pressure excess. In addition one defines the two components $\beta_\|,\beta_\perp$ of the plasma-$\beta$, with $\beta=2\mu_0NT/B^2$, as usual. Then the simplest {\it necessary} linear condition for the mirror instability \cite{hase1969} to be excited can be written in the extraordinarily simple form $A\beta_\perp>1$ . More sophisticated approaches including finite Larmor radius effects \cite{hall1979}, finite electron temperatures \cite{pantel1995,pokh2000}, kinetic corrections and non-maxwellian distributions \cite{pokh2000} and so on, simply modify this condition by assigning slightly different meanings (and in some cases also functional dependencies on other plasma parameters) to the anisotropy $A$. Surprisingly, even though the theory of the mirror mode is half a century old, its mechanism has still not been satisfactorily clarified. One of the reason is that the mirror mode competes with the electromagnetic ion cyclotron instability (EMCI) which grows under exactly the same and apparently even less restrictive conditions. The necessary condition for its excitation is $A>0$. One would thus expect that the mirror instability looses in growing against the EMCI under realistic conditions.

Observations do however show that the mirror instability can exist for a wide range of conditions in many places in space plasmas. It has been observed in the magnetosphere, the magnetosheath \cite{tsur1982}, the solar wind \cite{horb2004}, in cometary interactions \cite{russell1999} with the solar wind, and in the downstream vicinity \cite{czay2001} of Earth's bow shock wave. Most intensely it has been investigated using measurements in the magnetosheath \cite{lucek1999,const2003,soucek2008,genot2006} (for a review of the early observations and their interpretation see \cite{schwartz1996}). Moreover, it has been reproduced under specially tailored conditions as well in numerical simulations \cite[and others]{mckean1992,calif2008}. In the observations the mirror mode cannot be detected in its infinitesimal linear state of growth. It is always observed in a well developed though evolving time dependent not necessarily final state when it already has reached large amplitude and forms deep magnetic holes that are filled with hot plasma. Such a state must, however, have emerged from the linear state of the instability, and this implies that the instability has survived the competition with the EMCI. A stationary nonlinear solution has been constructed in \cite{const2002}. Physical arguments for its nonlinear evolution have been provided in \cite{kiv1996,treu2004,pokh2000} from different points of view.  

Within the limited range of nonlinear simulations \cite{mckean1992,calif2008} it has been suggested that mirror modes in the fluid picture form long extended bottles. The stationary analytical calculation \cite{const2002} does not necessarily confirm this conclusion. The more recent hybrid simulations \cite{calif2008} on the other hand show, in agreement with the linear and nonlinear theories \cite{kuz2007,pokh2000}, that mirror modes are short structures. Their linear growth rates have been shown to maximize in high-$\beta$ conditions at transverse wavelengths $\lambda_\perp$ of the order of the ion inertial length $\lambda_i=c/\omega_{pi}$, depending however on the environmental conditions such that slightly longer wavelength may also grow when the short waves cannot evolve \cite{pokh2000}. Observationally the mirror mode behaves surprisingly. Under very similar conditions in the magnetosheath, for instance, it may evolve into short perpendicular and long parallel wavelength magnetic holes or also into closely packed magnetic walls. The hybrid simulations \cite{calif2008} suggest that small change sin the parameters might produce magnetic holes with no susceptible enhancement of the surrounding magnetic field and also holes with steep magnetic overshoots at the boundaries. Such solutions may apply to the two conditions observed in the magnetosheath when the mirror mode evolves. It should, however be noted that the magnetosheath plasma is bounded from two sides, the bow shock in the upstream solar direction and the magnetopause in downstream direction, and the plasma where the mirror holes are embedded is flowing. The effect of these boundaries on the evolution of the holes is not known and has not been included neither in theory nor in the simulations yet.

The simpler problem is the competition between the EMCI and mirror modes. The competition between the two modes poses an unresolved problem which is physically minor but requires to be understood. In the magnetosheath both instabilities could grow. In an attempt to circumvent the problem \cite{gary1992} the reason for the presence of the mirror mode has been attributed to the presence of a substantial fraction of He ions, which indeed slightly reduces the growth of the EMCI. On the other hand, the mirror mode has also been seen in the absence of He ions. However, there is no real need to invoke any complications like the presence of heavy ions or strong modification by nonlinear theory for finding the parameter ranges of the mirror and EMC instabilities. Simple linear theory suffices for this purpose.  

\begin{figure}[t!]
\centerline{\includegraphics[width=0.5\textwidth,height=0.5\textwidth,clip=]{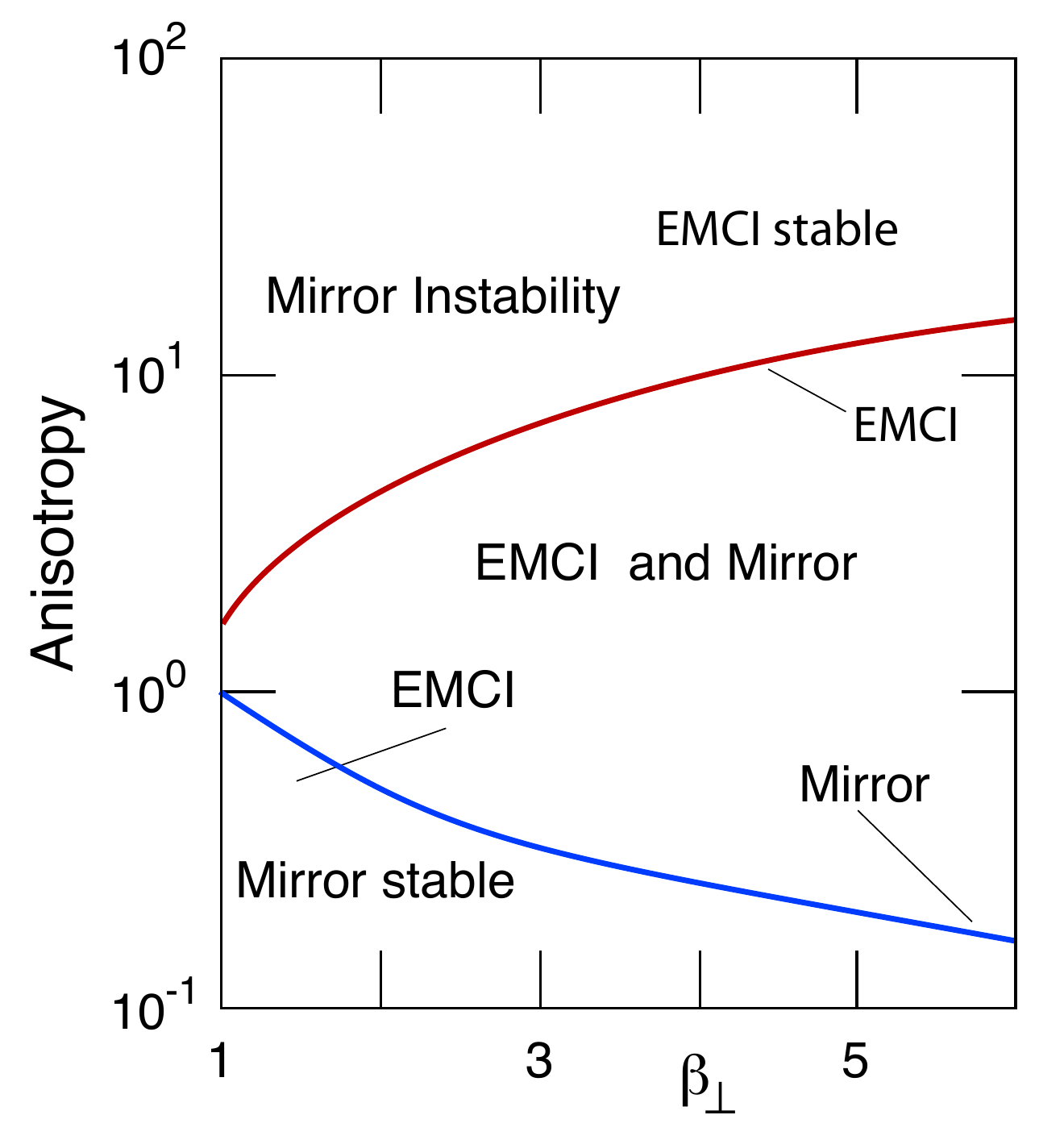} }
\caption[ ]
{\footnotesize  The existence ranges  and thresholds of the EMCI (red) and mirror (blue) instabilities in the $(A,\beta_\perp)$-plane. The EMCI grows below the red curve, the mirror mode grows above the blue curve. At small anisotropies the mirror mode is stable and only the EMCI can be excited. In the middle range both instabilities coexist, but the one with the larger growth rate will win as suggested in \cite{gary1992}. The mirror instability clearly dominates at large anisotropies where the EMCI turns out to be stable. }\label{chap5-fig-mirror}
\vspace{-0.3cm}\end{figure}
Both modes require that the anisotropy $A=P_{i\perp}/P_{i\|}-1>0$. However, the EMCI is a resonant instability which depends on the presence of ions with parallel energy ${\cal E}_{i\|}>m_iV_A^2/2=B^2/2\mu_0N$. This last condition is crucial. It has been omitted in all attempts to solve the problem where the focus was mostly numerical, considering particular parameter combinations in the calculation of the growth rates of the two competing instabilities and thus restrict to the domain where both are can exist simultaneously. 

For a bi-Maxwellian ion distribution with parallel and perpendicular temperatures $T_\|,T_\perp$, expressing ${\cal E}_{i\|}$ and the number of resonant particles $N_{i,{\rm res}}$ through $A$ and $\beta_\perp$, i.e. eliminating $\beta_\|$ in favour of $\beta_\perp$ by using the definition of $A$, yields that
\begin{equation}
\frac{{\cal E}_{i\|}}{T_{i\|}}=3\sqrt{2}-\left(\frac{A+1}{\beta_\perp}\right)^\frac{1}{2}>\frac{A+1}{\beta_\perp}
\end{equation}
where the right part of the inequality is just the energy condition on the  EMCI rewritten in terms of $A$ and $\beta_\perp$. One first observes that, for retaining positive resonant particle energies, one requires $A<18\beta_\perp-1$. Solving for the conditions on the right restricts this limit even further to
\begin{equation}
A<\frac{\beta_\perp}{4}\left[\left(1+12\sqrt{2}\right)^\frac{1}{2}-1\right]^2-1
\end{equation}
The EMCI thus exists only in the anisotropy range  $0<A<2.62\,\beta_\perp-1$ and for $\beta_\perp>0.38>\beta_\|$, which is a consequence of the requirement that sufficiently many resonant particles must be present in order to drive the EMC instability unstable. The condition on $\beta_\perp$ is unproblematic for the mirror instability as the mirror instability refers to high-$\beta$ plasma only. 

The condition for the mirror instability, which is an instability that resonates with the abundant low parallel energy ($v_\|\sim 0$) ions \cite{south1993,pokh2000}, is simply $A\beta_\perp>1$, subject to the limitations imposed above. The boundaries of the regions in $(A,\beta_\perp)$-space where these conditions are satisfied for the mirror and EMC instabilities are shown in Figure \ref{chap5-fig-mirror}. There is a large domain between the two curves where both instabilities coexist and where the one with the larger growth rate will win as was suggested in \cite{gary1992}. However, at large anisotropies above the EMCI curve solely the mirror instability can be excited. With increasing $\beta_\perp$ the anisotropy must become large as well in order still to excite the mirror mode. This is clear from the condition that the number of high energy resonant particles grows when $\beta_\perp$ increases. On the other hand, at small anisotropies the mirror instability is stable while the EMCI can grow for all $\beta_\perp$. 
\begin{figure}[t!]
\centerline{\includegraphics[width=0.5\textwidth,height=0.5\textwidth,clip=]{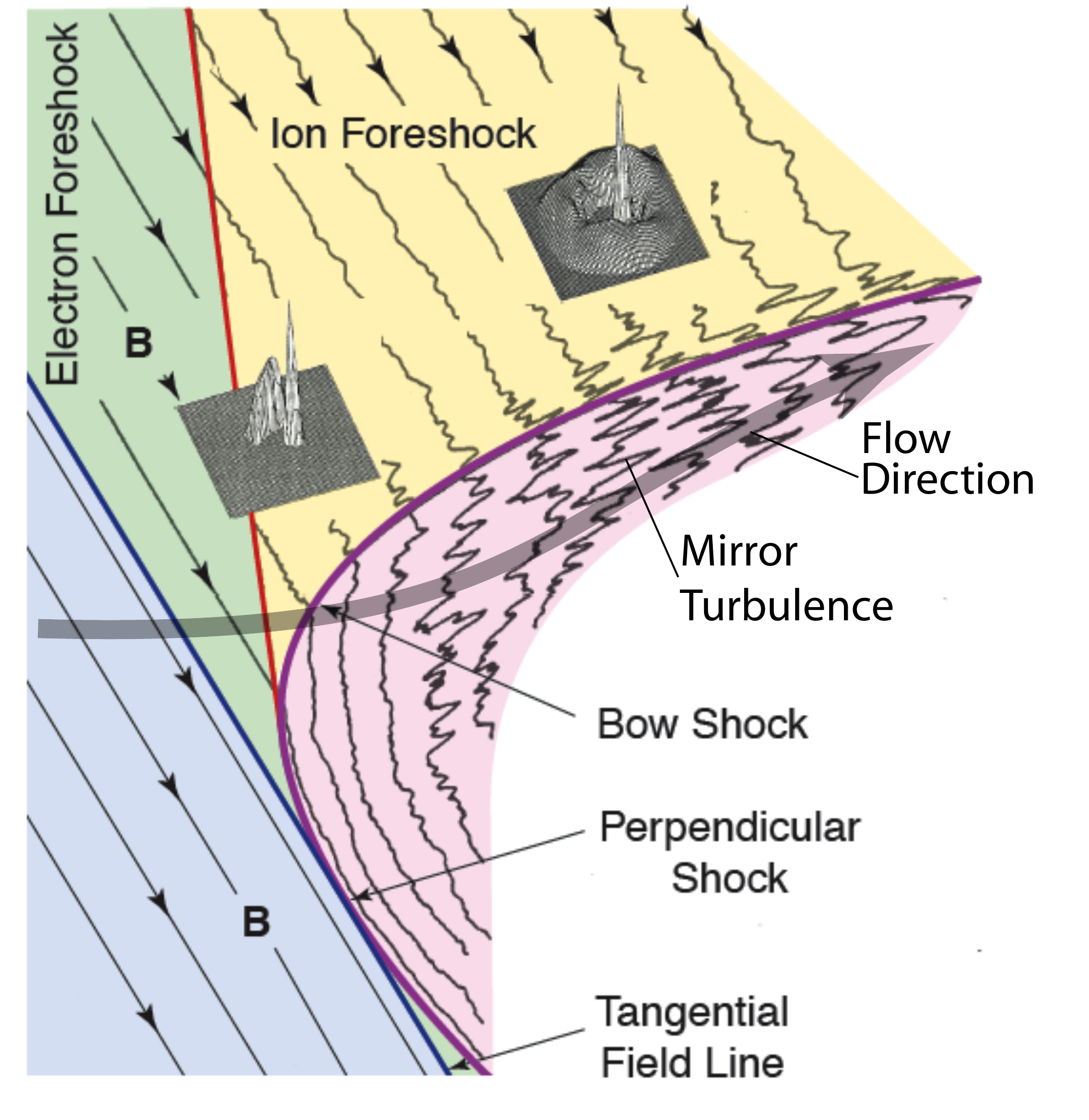} }
\caption[ ]
{\footnotesize  The bow shock-magnetosheath system. Shown is the flow direction and the evolution of downstream mostly mirror mode turbulence in the magnetosheath behind the bow shock. The important information in this figure aside from the geometry of the structure is that the flow is mostly along the magnetosheath thus increasing the e-folding time of the mirror instability. The inserts in the foreshock show the reflected ions in phase space which when passing ulitmately across the shock to downstream both contribute to anisotropy and to excitation of EMCI waves just behind the shock, while the fluid anisotropy that drives the mirror mode has first to evolve along the flow.}\label{chap5-fig-fsbc}
\vspace{-0.3cm}\end{figure}

Anisotropies of the order of $A\sim 10$ are not unusual in the magnetosheath in which case, for moderate $\beta_\perp$, the external conditions may well lie in the range of suppressed EMCI thus favouring the evolution of the mirror instability even in the absence of a substantial fraction of heavy ions. Kinetic \cite{pokh2000} and nonlinear effects do indeed slightly modify the marginal stability boundary and growth rates of the mirror instability. In general they lift the lower bound of the entire mirror unstable region in Figure \ref{chap5-fig-mirror} upward displacing parts of this curve into the domain that is shared by the mirror and EMC instabilites and thus decreasing the unstable mirror range at {\it low} anisotropies $A<1$. The upper mirror unstable range remains less affected (or even unaffected) by these effects. In that range, even for low growth rates $\gamma\ll\omega_{ci}$ of the mirror instability one may expect that the mirror mode can reach fairly large amplitudes in a typical convection time across, say, the magnetosheath. 

For an estimate we consider magnetosheath conditions. The angular ion cyclotron frequency in the magnetosheath is of the order of $\omega_{ci} \sim 10\,{\rm s}^{-1}$. With a realistic assumption on $\gamma\sim 0.01\omega_{ci}$, flow velocity of $V\sim 100$\,km/s, and the  width of the magnetosheath being of the order of $\sim 2$\,R$_{\rm E}\gtrsim 10^4$\,km (see Figure \ref{chap5-fig-fsbc}), a transverse convective crossing of the magnetosheath just corresponds to one e-folding time of 100 s for a magnetic mirror hole, if the hole would form right behind the bow shock. However, the anisotropy must first evolve from the bow shock along the convective path to reach into the mirror unstable domain. Behind the shock one has $\beta_\perp\sim1-2$. Hence, an anisotropy of $A\sim 5-10$ is needed for the mirror mode to not compete with the EMCI. The latter we know is excited right behind the shock \cite{sckopke1995}. Anisotropies of this magnitude evolve roughly at about half the distance between the shock and the magnetopause. Interestingly enough, it has been found in the observations that at about this distance the state of low frequency wave excitation in the magnetosheath changes about abruptly \cite{hill1995}, indeed. This apparently reduces the accessible e-folding time. However, the flow is not crossing the magnetosheath straight from the shock to hit the magnetopause. At the contrary, it is deflected and is turned around the magnetopause, and observationally the strongest mirror activity is found close to the magnetopause at either the flanks or at higher latitudes. Thus the flow moves roughly a distance of 5-8 R$_{\rm E}\gtrsim 6\times 10^4$\,km until the mirror modes have evolved to the observed amplitudes. This yields a growth time of $\tau\sim600$ s or six e-folding times corresponding to a factor e$^6\sim 400$ in the magnetic mirror amplitude. The observations show that mirror modes in their evolved state, when they are observed in the magnetosheath, have hole amplitudes $\delta B-$50-80\% (less than) the surrounding ambient field value. The latter is of the order of $B\sim 30$ nT, which suggests that the initial infinitesimal mirror wave amplitude at about 1 R$_{\rm E}$ behind the bow shock was of the order of $|\delta B|\sim 0.1-0.3$ nT, which at these low mirror frequencies $\omega\sim 0$ is in the measurement noise. This estimate is consistent with our assumptions on the growth of mirrors. 

Nevertheless, the nonlinear state of the mirror mode remains to be an unresolved problem, as is the basic physics of the mirror process. We have elucidated the related problems in an earlier paper \cite{treu2004}. In sharpening the argument that has been presented there, we note that, in contrast to most other plasma instabilities, the mirror mode is an exception as the mirrors are no real waves. This contrasts for instance its complement, the firehose mode which in an anisotropic plasma is a travelling Alfv\'en wave, for all its amplitudes, even the largest ones. The mirror mode instead is a plasma structure. In its infinitesimal initial state this does not pose a problem, but once it starts growing it is in conflict with thermodynamics as it by itself creates structure, forming a series of magnetic holes in the hot plasma with the holes being of the scale of the ion inertial length containing grossly unmagnetised ions and magnetised electrons. Structure formation is, however, related to phase transitions. Hence the formation of mirror holes in hot collisionless plasma resembles an internal phase transition of which the plasma is capable. Indeed, the condition for instability can be rewritten in the form
\begin{equation}
B^2<2\mu_0NT_\perp A\equiv B_{crit}^2
\end{equation}
where we introduced a critical magnetic field $B_{crit}=\sqrt{2\mu_0NT_\perp}$. If the magnetic field in an anisotropic plasma drops below this critical value, phase transition sets on, and the plasma starts developing structure, breaking off into magnetic hole which are surrounded by regions of `normal' plasma state. This argument resembles the Meissner effect in superconductivity while it here is applied to the ideally conducting (collisionless) state of a high temperature plasma. In the same terminology the formation of structure on the scale $\lambda_i\ll L$, where $L$ is the macroscale of the plasma, then means that this phase transition is similar to a phase transition of a superconductor of the second kind. Of course, this is just a similarity or an analogy and not the same physics, as the mirror mode is not in a quantum state. It shows, however, that the physics of the mirror mode in high temperature plasma contains some most interesting problems which are still badly understood. 

{
{\small{\bf\textsf{Acknowledgements.}} This research is part of a Visiting Scientist Programme at ISSI, Bern. It has also benefitted from a Gay-Lussac-Humboldt award of the French Government. The author acknowledges discussions with Andr\'e Balogh and Johannes Geiss, directors at ISSI, as well as the hospitality of the ISSI directors, Roger-Maurice Bonnet, Lennart Bengtsson and Rudolf von Steiger and the ISSI staff.}}

\end{document}